\documentclass{optica-article}
\journal{opticajournal} 

\articletype{Research Article}

\usepackage{lineno}

\usepackage{amsmath,amsfonts}
\usepackage{graphicx}
\usepackage{physics}
\usepackage{float}

\newcommand{\A}{\mathbf{A}}
\newcommand{\B}{\mathbf{B}}
\newcommand{\C}{\mathbf{C}}
\newcommand{\D}{\mathbf{D}}
\newcommand{\Binv}{\mathbf{B}^{-1}}
\newcommand{\ro}{\mathbf{r}_{1}}
\newcommand{\ri}{\mathbf{r}_{2}}
\newcommand{\rotilde}{\mathbf{\tilde{r}}_{1} + \mathbf{r}_{1M}}
\DeclareMathAlphabet\mathbfcal{OMS}{cmsy}{b}{n}

\begin{document}


\title{A Generalized Expression for Accelerating Beamlet Decomposition Simulations}
\author{Jaren N. Ashcraft\authormark{1,*} Ewan S. Douglas\authormark{2}, Ramya Anche\authormark{2}, Brandon D. Dube\authormark{3}, Kevin Z. Derby\authormark{1}, Lars Furenlid\authormark{1,4}, Maggie Kautz\authormark{1}, Daewook Kim\authormark{2,5}, Kian Milani\authormark{1}, A J Eldorado Riggs\authormark{3}} 

\noindent \authormark{1}{James C. Wyant College of Optical Sciences, University of Arizona, Meinel Building 1630 E. University Blvd., Tucson, AZ. 85721, USA} \\
\authormark{2}{Department of Astronomy and Steward Observatory, University of Arizona, 933 N. Cherry Ave., Tucson, AZ 85721, USA} \\
\authormark{3}{Jet Propulsion Laboratory, California Institute of Technology, 4800 Oak Grove Drive, Pasadena, CA 91109, USA} \\
\authormark{4}{College of Medicine, Department of Medical Imaging, P.O. Box 245067 Tucson, Arizona 85724-5067, United States} \\
\authormark{5}{Large Binocular Telescope Observatory, University Of Arizona, 933 N. Cherry Ave. Tucson, AZ 85721, USA} \\

\email{\authormark{*}jashcraft@arizona.edu} 


\begin{abstract*} 
Paraxial diffraction modeling based on the Fourier transform has seen widespread implementation for simulating the response of a diffraction-limited optical system. For systems where the paraxial assumption is not sufficient, a class of algorithms has been developed that employs hybrid propagation physics to compute the propagation of an elementary beamlet along geometric ray paths. These ``beamlet decomposition" algorithms include the well-known Gaussian Beamlet Decomposition (GBD) algorithm, of which several variants have been created. To increase the computational efficiency of the GBD algorithm, we derive an alternative expression of the technique that utilizes the analytical propagation of beamlets to tilted planes. We then use this accelerated algorithm to conduct a parameter-space search to find the optimal combination of free parameters in GBD to construct the analytical Airy function. The experiment is conducted on a consumer-grade CPU, and a high-performance GPU, where the new algorithm is 34 times faster than the previously published algorithm on CPUs, and 67,513 times faster on GPUs.
\end{abstract*}

\section{Introduction}
\label{sec:intro}
Physical optics models are integral to the design and tolerancing of diffraction-limited optical systems. Traditional diffraction theory derived from the Huygens-Fresnel principle enforces the assumption that the optical system is scalar and paraxial in order to express the propagated field in terms of the Fourier Transform. To support plane-to-plane propagation, we approximate the propagation of an optical field ($E(r)$) as a projection onto a parabolic phase kernel, resulting in the Fresnel diffraction integral given in Eq. \ref{eq:fresnel},

\begin{equation}
	E_{2}(\mathbf{r}_{2},z) = \frac{e^{ikz}}{i\lambda z} e^{\frac{ik}{2z}|\mathbf{r}_{2}|^{2}}\iint_{-\infty}^{\infty} E_{1}(\mathbf{r}_{1},0) e^{\frac{ik}{2z}|\mathbf{r}_{1}|^{2}} e^{\frac{-i2\pi}{\lambda z} (\mathbf{r}_{1} \cdot \mathbf{r}_{2})}d^{2}\mathbf{r}_{1}.
	\label{eq:fresnel}
\end{equation}

Here $\mathbf{r}_{2}$ is the radial coordinate at the plane of evaluation at distance $z$, $\mathbf{r}_{1}$ is the radial coordinate at the plane $z=0$ where the propagation begins, $d^{2}\mathbf{r}_{1}$ is the differential for the two dimensional integration, $k$ is the wavenumber of light and $\lambda$ is the wavelength. Eq. \ref{eq:fresnel} is expressed such that the Fourier kernel (rightmost exponential term in Eq. \ref{eq:fresnel}) is separated from the parabolic phase term over the source coordinates $\mathbf{r}_{1}$. In the limit where $z \gg |\mathbf{r}_{1}|$, the parabolic phasor in the integrand disappears and we are left with the Fraunhofer diffraction integral in Eq. \ref{eq:fraunhofer},

\begin{equation}
	E_{2}(\mathbf{r}_{2},z) = \frac{e^{ikz}}{i\lambda z} e^{\frac{ik}{2z}|\mathbf{r}_{2}|^{2}}\iint_{-\infty}^{\infty} E_{1}(\mathbf{r}_{1},0) e^{\frac{-i2\pi}{\lambda z} (\mathbf{r}_{1} \cdot \mathbf{r}_{2})}d^{2}\mathbf{r}_{1},
	\label{eq:fraunhofer}
\end{equation}

which represents a simple Fourier transform of the field $E_{1}$. Equations \ref{eq:fresnel} and \ref{eq:fraunhofer} form the cornerstone of modern scalar diffraction modeling and have proven to be an excellent tool for the modern era in open-source diffraction modeling packages\cite{2016ascl.soft02018P,Dube2022,Dube2019,HCIPYdocs}. However, these formulations enforce that the optical system is paraxial and neglects ``ray aberration" effects like wavefront and pupil aberration. These effects are captured best by ray trace models of optical systems, which ignore diffraction. To capture these effects simultaneously, one can employ a ``hybrid" approach to propagation physics where several models are linked to form a more physical simulation. This can be done by computing the scalar wavefront error with a ray trace to the exit pupil and then using the diffraction integrals above to propagate to the image plane. However, linking multiple models simultaneously is error-prone, requiring considerable effort to ensure the propagation physics are appropriately synched. Furthermore, the scalar and paraxial assumption is insufficient to model some optical systems. One example is high-numerical aperture systems, like microscope objectives \cite{mansuripur_distribution_1986} or highly aspheric mirrors\cite{krist_practical_2010}.

\begin{figure}
	\centering
	\includegraphics[width=0.8\textwidth]{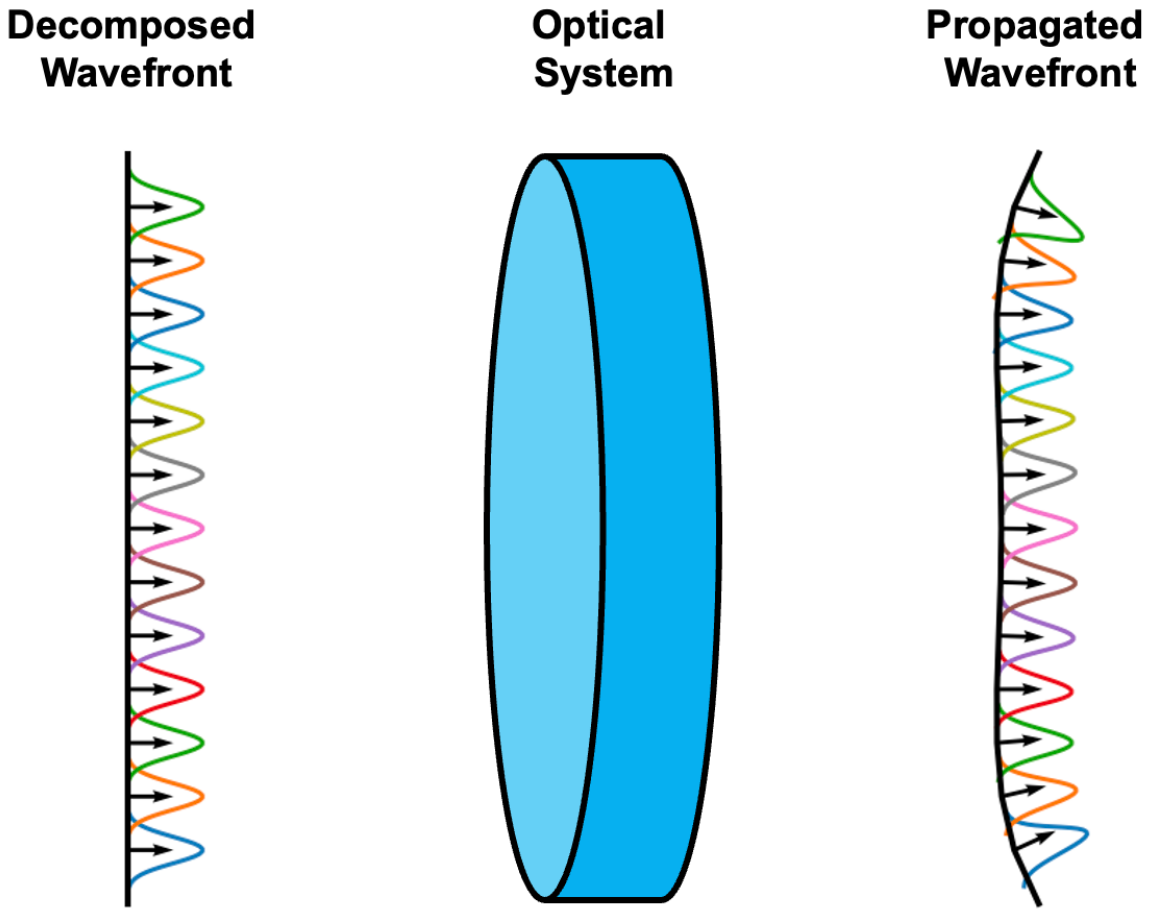}
	\caption{Illustration of Gaussian Beamlet Decomposition. An input wavefront (left) is decomposed into a sum of Gaussian beams. They are propagated through an optical system, and the resultant aberrated wavefront is represented by a change in optical path of each of the beamlets (right).}
	\label{fig:gbd_principle}
\end{figure}

A more elegant approach would be to perform the diffraction and ray trace simulations simultaneously. To do so, we consider the Gaussian Beamlet Decomposition technique (GBD)\cite{Harvey15,Greynolds14,Greynolds86,Worku17,Worku:18,Worku:20}. GBD is a method of physical optics propagation where the propagated field is simulated as a finite summation of Gaussian beams (as shown in Eq. \ref{eq:sum_gaussian}) propagated normal to the local wavefront (as shown in Fig. \ref{fig:gbd_principle})

\begin{equation}
	E_{2}(\mathbf{r}_{2},z) = \sum_{j = 0}^{N} \frac{E_{o,j}}{q(z)_{j}} e^{(\frac{-ik|\mathbf{r}_{2}-\mathbf{r}_{j}|^{2}}{2q(z)_{j}})},
	\label{eq:sum_gaussian}
\end{equation}

Where the argument of the sum shows the field for the $j-th$ Gaussian beam shifted by $\mathbf{r}_{j}$ in the entrance pupil and whose complex beam parameter is $q(z)_{j}$. The complex beam parameter is related to the waist radius ($w(z)$)  and wavefront radius of curvature ($R(z)$) of a Gaussian beam by Eq. \ref{eq:complexq}

\begin{equation}
	q(z)^{-1} = \frac{1}{R(z)}+i\frac{\lambda}{\pi w(z)^2}.
	\label{eq:complexq}
\end{equation}

The field propagation in Eq. \ref{eq:sum_gaussian} is valid as long as the beamlets that constructed it represent the initial field well. Traditional GBD employs solely the fundamental Gaussian mode, which is incapable of representing the sharp edges seen in conventional imaging systems well. However, GBD has recently seen substantive development by Worku and Gross, which generalized the propagation of a Gaussian Beam to include curved\cite{Worku17}, truncated\cite{Worku19}, polarized\cite{Worku:18}, and spectrally chirped\cite{Worku:20} Gaussian beamlets to use in decomposition to increase the accuracy of the simulation. This modified GBD (MGBD) approach illustrates the flexibility of beamlet decomposition for high-fidelity integrated modeling of ray aberration, diffraction, and polarization. In a prior work\cite{ashcraft_inreview}, we derived an implementation of MGBD and implemented it in an open-source propagation package\cite{Ashcraft_2023_poke} to determine if GBD was suitable for astronomical high-contrast imaging simulations. That study found the runtime required to conduct highly-sampled simulations was a major limitation. This was due to MGBD requiring the propagation of every Gaussian beamlet to a single point in the plane where the propagated field would be evaluated. Furthermore, since the decomposition of a wavefront into a discrete set of Gaussian beams does not have a unique solution, we need to explore the free parameters available in MGBD simulations in order to determine what combination delivers the most accurate result. This requires iterating on the different parameters, which can take a considerable amount of time. 

In this study, we derive an alternative expression for MGBD inspired by the work of Weber configuring the general Collins integral for misaligned optical elements \cite{Weber06} that enables the propagation of a Gaussian beam to a tilted plane instead of a single point. We then validate the accuracy of the proposed ``plane-evaluation" method against the ``point-evaluation" method we previously derived\cite{ashcraft_inreview,Ashcraft_2023_poke}, and perform a runtime comparison on CPUs and GPUs. In Section \ref{sec:methods}, we review the point-evaluation method derived for solutions of the Collins integral. In Section \ref{sec:algorithm}, we describe Weber's expression of the Collins integral for misaligned elements and our modification to their derivation to support efficient MGBD without loss of generality. In Section \ref{sec:results}, we perform self-consistency tests between the plane-evaluation algorithm, Fresnel diffraction, and the point-evaluation algorithm to demonstrate the accuracy of the propagation physics. In Section \ref{sec:runtime_compare} we quantify the runtime gains achievable through using the plane-evaluation algorithm over the point-evaluation algorithm. In Section \ref{sec:paramsearch}, we use the new algorithm to perform a parameter space search of the optimal sampling conditions for the various degrees of freedom used in GBD, including overlap factor and number of beamlets used.


\section{Methods}
\label{sec:methods}

\subsection{The General Collins Integral}
The general Collins integral \cite{Collins:70} is a reformulation of second-order diffraction theory that relates the diffraction from one plane orthogonal to the propagation axis to another via a nonorthogonal ray transfer matrix whose elements are matrices $\A,\B,\C,\D$. The rotationally symmetric version is a direct analog to the Fresnel diffraction integral in Eq. \ref{eq:fresnel} and is expressed as,
\begin{equation}
	E_{2}(\mathbf{r}_{2}) = \frac{1}{\lambda B} e^{D|\mathbf{r}_{2}|^{2}} \iint_{-\infty}^{\infty} E_{1}(\mathbf{r}_{1}) exp[\frac{i\pi}{\lambda B}(A|\mathbf{r}_{1}|^{2} - 2(\mathbf{r}_{1} \cdot \mathbf{r}_{2}) + D|\mathbf{r}_{2}^{2}|)d^{2}\mathbf{r}_{1},
	\label{eq:ortho_collins}
\end{equation}
where $A, B,$ and $D$ are elements of the paraxial 2$\times$2 ray transfer matrix\cite{goodman17}. The general integral for nonorthogonal optical systems is given by \cite{Collins:70,Worku19} Eq. \ref{eq:collins},

\begin{equation}
	E_{2}(\mathbf{r}_{2}) = \frac{i\ n_{1}}{\lambda\ det(\mathbf{B})^{1/2}} exp(-ikl_{o}) \iint_{-\infty}^{\infty} E(\mathbf{r}_{1}) exp(-ikl_{1}) d^{2}\mathbf{r}_{1},
	\label{eq:collins}
\end{equation}

where $n_{1},n_{2}$ are the refractive indices of the incident and exiting media, $k$ is the wave number, $l_{o}$ is the axial propagation distance, and $l_{1}$ represents the transformation of the field by the ABCD optical system, which is given by Eq. \ref{eq:eikonal1},

\begin{equation}
	l_{1} = \frac{1}{2}
	\begin{pmatrix}
		\mathbf{r}_{1} \\
		\mathbf{r}_{2} \\
	\end{pmatrix}^{T}
	\begin{pmatrix}
		n_{1}\Binv \A & -n_{1}\Binv \\
		n_{2}(\C - \D \Binv \A) & \D \Binv \\
	\end{pmatrix}
	\begin{pmatrix}
		\mathbf{r}_{1} \\
		\mathbf{r}_{2} \\
	\end{pmatrix},
	\label{eq:eikonal1}
\end{equation}

where the superscript $T$ denotes the transpose and $\ro,\ri$ are the radial coordinates in the transverse plane before and after propagation, respectively. In free space, the refractive indices of the incident and exiting media are equal to unity $n_{1} = n_{2} = 1$. Furthermore, since the ray transfer matrix is symplectic, it obeys the following relations:

\begin{equation}
	\B\A^{T} = \A\B^{T}
	\label{eq:symp1}
\end{equation}
\begin{equation}
	\B^{T}\D = \D^{T}\B
	\label{eq:symp2}
\end{equation}
\begin{equation}
	\D\C^{T} = \C\D^{T}
	\label{eq:symp3}
\end{equation}
\begin{equation}
	\A^{T}\C = \C^{T}\A
	\label{eq:symp4}
\end{equation}
\begin{equation}
	\A\D^{T} - \B\C^{T} = \mathbf{I},
	\label{eq:symp5}
\end{equation}

where $\mathbf{I}$ is the identity matrix. Nazarathy showed that Eq. \ref{eq:collins} can alternatively be expressed as Eq. \ref{eq:reduced_collins} \cite{Siegman_1986}.

\begin{equation}
	E_{2}(\mathbf{r}_{2}) = K \iint_{-\infty}^{\infty} E(\mathbf{r}_{1}) exp(\frac{-ik}{2}[\bra{\ri}\D\Binv\ket{\ri} + \bra{\ro}\Binv\A\ket{\ro} - 2\bra{\ro}\Binv\ket{\ri}]) d^{2}\mathbf{r}_{1},
	\label{eq:reduced_collins}
\end{equation}

Where $K$ are the constants in front of the integral in Eq. \ref{eq:collins}, and the "Bra-Ket" notation is chosen to be consistent with Weber's notation \cite{Weber06}. Since the vectors and matrices are real-valued, $\bra{\mathbf{v}}$ is equivalent to $\ket{\mathbf{v}}^{T}$. Equation \ref{eq:reduced_collins} has been solved by several authors to determine the propagation laws for Hermite\cite{CAI2002139}, Laguerre\cite{Mei_2005}, truncated\cite{Worku19}, and pulsed\cite{Worku:20} Gaussian beams. All of these solutions can be employed in beamlet decomposition for diffraction simulation.

\subsection{The Point-Evaluation Approach}
\begin{figure}
	\centering
	\includegraphics[width=\textwidth]{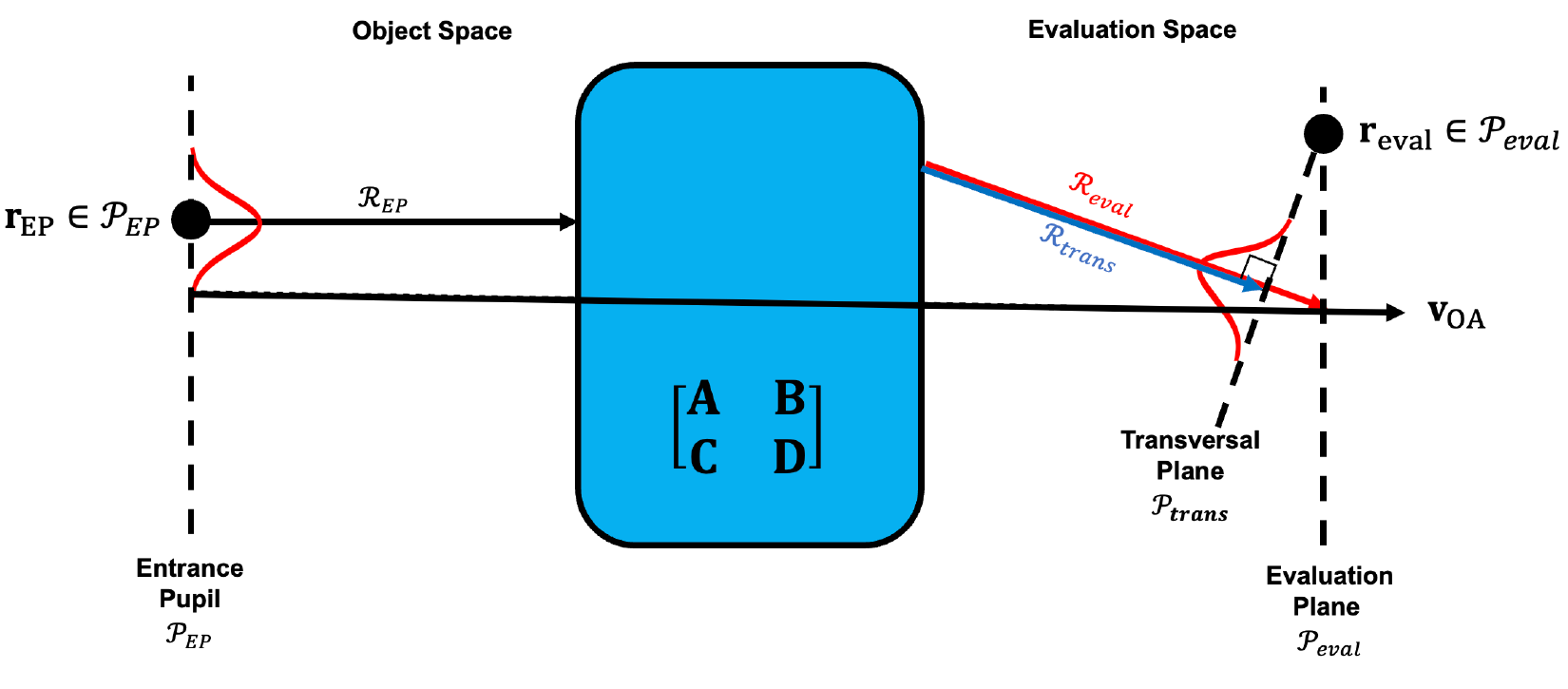}
	\caption{Schematic of the ``traditional" GBD algorithm referred to in this work. The decomposition is referred to at the entrance pupil of the optical system $\mathbfcal{P}_{EP}$, where rays that emanate normal to the center of a beamlet ($\mathbfcal{R}_{EP}$) are propagated through an optical system to the plane where we desire to evaluate the field $\mathbfcal{R}_{eval}$. In order to perform the field evaluation, the rays must be transformed to the transversal plane $\mathbfcal{P}_{trans}$ and the Gaussian field evaluated at $\mathbf{r}_{eval}$ expressed in the coordinates of the transversal plane. The procedure to do so is described in \cite{ashcraft_inreview}.}
	\label{fig:traditional_gbd}
\end{figure}

To perform GBD for an optical system, we decompose the wavefront in the optical system's entrance pupil into a sum of Gaussian beams. The result of this decomposition is a set of beamlets whose propagation are described by 5 parabasal rays for each beamlet \cite{Harvey15, Greynolds86, Worku19}. This list of rays in the entrance pupil are represented by $\mathbfcal{R}_{EP}$. We use these rays in a GBD simulation by carrying out the algorithm below for each point ($\mathbf{r}_{eval}$) for which the field is evaluated:

\begin{enumerate}
	\item Propagate the rays $\mathbfcal{R}_{EP}$ to the plane where the diffracted field will be evaluated $\mathbfcal{P}_{eval}$ with ray tracing, resulting in a new collection of rays $\mathbfcal{R}_{eval}$. A member of $\mathbfcal{R}_{EP}$ is shown as a black arrow in Fig. \ref{fig:traditional_gbd}.
	\item Determine the propagation distance from the rays at the evaluation plane $\mathbfcal{R}_{eval}$ (shown in red on Fig. \ref{fig:traditional_gbd}) to the plane normal to the rays and intersecting the point at which the field will be evaluated $\mathbf{r}_{eval}$, called the transversal plane $\mathbfcal{P}_{trans}$.
	\item Propagate $\mathbfcal{R}_{eval}$ to $\mathbfcal{P}_{trans}$ and rotate them into the local coordinate system of the transversal plane, resulting in a new collection of rays $\mathbfcal{R}_{trans}$ (shown in blue on Fig. \ref{fig:traditional_gbd}). 
	\item Compute the differential ABCD matrix from the propagation of $\mathbfcal{R}_{EP}$ to $\mathbfcal{R}_{trans}$ using differential ray tracing. 
	\item Using the ABCD matrix and $\mathbfcal{R}_{trans}$, evaluate the Gaussian field at $\mathbf{r}_{eval}$.
	\item Repeat steps 1-5 for every beamlet used in the decomposition of the entrance pupil wavefront, and coherently sum them at $\mathbf{r}_{eval}$.
\end{enumerate}

The precise mathematics of this procedure are illustrated in \cite{ashcraft_inreview}, but the simplification offered above reveals the inefficiency we wish to address. A substantial amount of computation is done in steps 2-4, where the traced rays are propagated from $\mathbfcal{P}_{eval}$ to $\mathbfcal{P}_{trans}$ for every $\mathbf{r}_{eval}$. It would be optimal to instead evaluate the contribution of a beamlet at the plane of points $\mathbfcal{P}_{eval}$ from the outset since we are frequently interested in the distribution of fields aligned to a plane (e.g., a detector in an imaging system). However, recall from Section 2.1 that the use of the Collins integral requires that the fields before and after propagation be defined on planes orthogonal to propagation\cite{Siegman_1986,Collins:70}. To solve for the influence of a beamlet on a plane tilted with respect to the propagation direction, we need an expression similar to the Collins integral to diffract to tilted planes.

\section{The Proposed Plane-Evaluation Approach}
\label{sec:algorithm}

\begin{figure}[H]
	\centering
	\includegraphics[width=\textwidth]{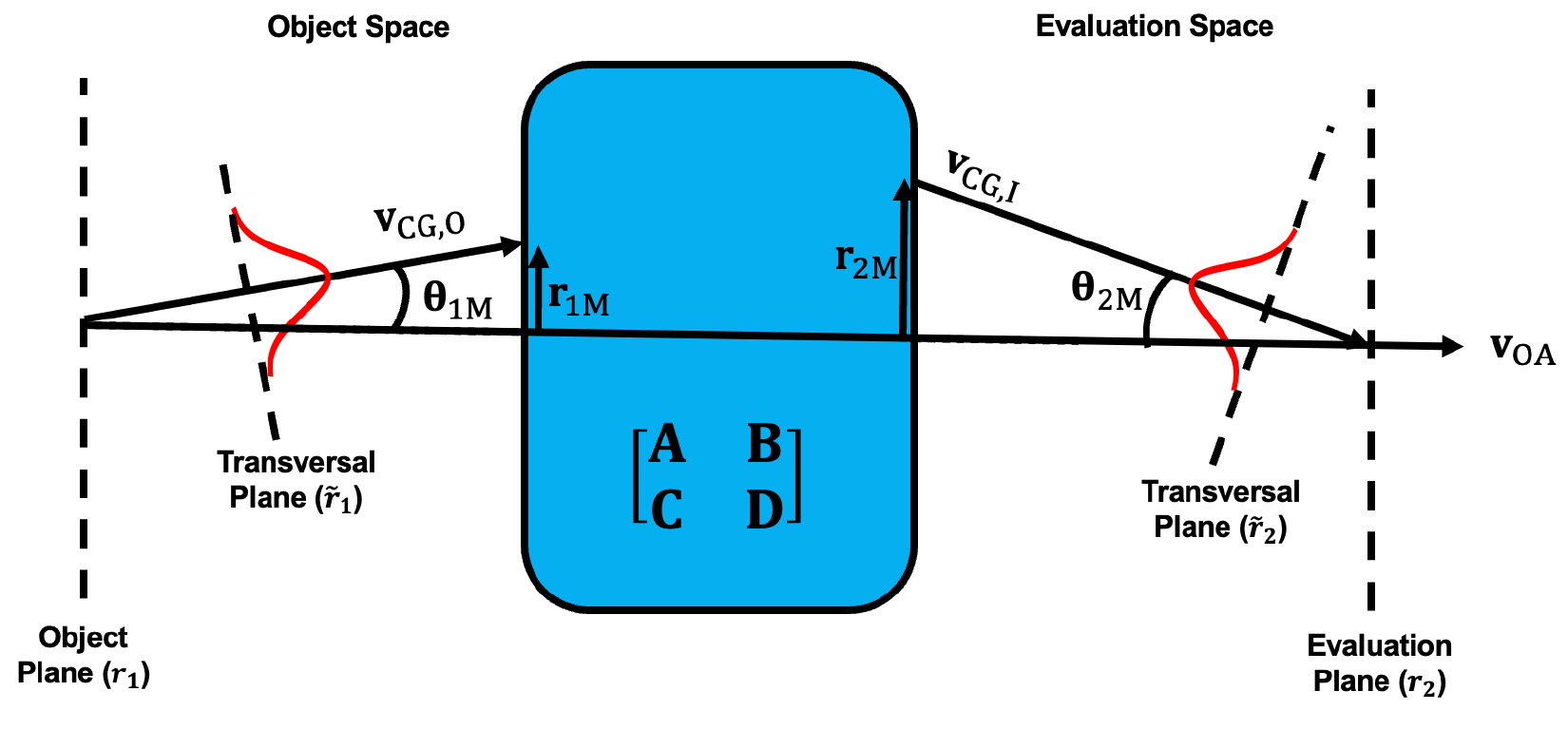}
	\caption{Illustration of Weber's misaligned optical system given by an arbitrary ray transfer matrix. Nominally, the Collins integral relates the field at the object plane (left side, dashed vertical line) to the evaluation plane (right side, dashed vertical line). Weber's formulation reconsiders the Collins integral propagation along the center of gravity vector in object space $\mathbf{v}_{CG,O}$ to the one in evaluation space $\mathbf{v}_{CG,I}$ by adding a misalignment vector in position ($\mathbf{r}_{1,2M}$) and angle ($\mathbf{\theta}_{1,2M}$) and evaluating the integral on the plane transverse to the centers of gravity $\mathbf{\tilde{r}}_{1,2}$.}
	\label{fig:weber_diagram}
\end{figure}

Weber \cite{Weber06} proposed an alternative expression of the general Collins integral to model wave propagation between misaligned optical elements. Weber's formulation generalizes the Collins integral to fields that propagate along a "center of gravity" vector $\mathbf{v}_{CG}$ that is generally not aligned to the optical axis $\mathbf{v}_{OA}$. A schematic of this propagation scheme is shown in Fig. \ref{fig:weber_diagram}. The misalignment vector is given by the difference between the center of gravity vector and the optical axis vector, as shown in Eq. \ref{eq:misalign_vec}.

\begin{equation}
	\mathbf{v}_{jM} = \mathbf{v}_{jCG} - \mathbf{v}_{jOA} = 
	\begin{pmatrix}
		\mathbf{r}_{jM} \\
		\mathbf{\theta}_{jM} \\
	\end{pmatrix}
	; j = 1,2
	\label{eq:misalign_vec}
\end{equation}

To generalize the propagation through the Collins integral, Weber performs a change of variables to integrate over the transversal plane in object space ($\mathbf{\tilde{r}}_{1}$) instead of the coordinates of the object plane ($\mathbf{r}_{1}$). Since the coordinates of the transversal plane are just a shift of the object plane coordinates by $\mathbf{r}_{1}$ and a rotation by $\mathbf{\theta}_{1}$, the transformed coordinates are given by Eq. \ref{eq:change_coords}

\begin{equation}
	\mathbf{v}_{1} = \tilde{\mathbf{v}}_{1} + \mathbf{v}_{1M}.
	\label{eq:change_coords}
\end{equation}

Weber performs this substitution in Eq. \ref{eq:reduced_collins} which results in the Collins integral that maps a beam propagating along a center of gravity vector in object space ($\mathbf{v}_{CG,O}$) to the coordinates of the evaluation plane ($\ri$), given by Eq. \ref{eq:collins_misalign}.

\begin{align} \label{eq:collins_misalign}
	E_{2}(\mathbf{r}_{2}) = & K \iint_{-\infty}^{\infty} E(\mathbf{\tilde{r}}_{1}) exp(\frac{-ik}{2}[\bra{\ri}\D\Binv\ket{\ri} &\notag
	\\ &+\bra{\rotilde}\Binv\A\ket{\rotilde}  \\
	& -2\bra{\rotilde}\Binv\ket{\ri}]) d^{2}\mathbf{\tilde{r}}_{1} &\notag
	\label{eq:collins_misalign}
\end{align}

Weber moves on to make the same substitution for the radial coordinate on the transversal plane in evaluation space. To see the proof for this, we direct the reader to their work \cite{Weber06}. The result is rather elegant, and given in Eq. \ref{eq:misalignphase} (Eq. 10 in Weber's paper). We have also reproduced part of the derivation in \hyperref[sec:appendixA]{Appendix A} of this work.

\begin{equation}
	E_{2}(\tilde{\ri}) = E_{2,aligned}(\tilde{\ri}) exp(\frac{-ik}{2}[\bra{\mathbf{r}_{1M}}\ket{\mathbf{\theta}_{1M}} - \bra{\mathbf{r}_{2M}}\ket{\mathbf{\theta}_{2M}}]),
	\label{eq:misalignphase}
\end{equation}

$E_{2,aligned}(\tilde{\ri})$ is the solution to the original Collins integral given by Eq. \ref{eq:collins}. This relation is robust because it generalizes the propagation of all solutions to the integral to be generally misaligned with respect to the optical axis. However, for beamlet decomposition techniques it is not immediately useful. Since the field in Eq. \ref{eq:misalignphase} is expressed in coordinates of the plane normal to a beamlet's propagation, all beamlets cannot be coherently summed at a common plane.

Thus, we desire an expression for the solution of the Collins integral where the field at all points of a common evaluation plane can be computed simultaneously for a single beamlet. To arrive at this expression, we return to Weber's derivation starting at Eq. \ref{eq:collins_misalign}. We expand the terms of Eq. \ref{eq:collins_misalign} and follow Weber's assertion that the terms linear in $\mathbf{\tilde{r}}_{1}$ vanish, so the resulting integral is given by Eq. \ref{eq:collins_deriv1}

\begin{align} \label{eq:collins_deriv1}
	E_{2}(\mathbf{r}_{2}) = K \iint_{-\infty}^{\infty} E(\mathbf{\tilde{r}}_{1}) exp(\frac{-ik}{2}[
	&\notag 
	\bra{\ri}\D\Binv\ket{\ri} \\
	+ & \notag \bra{\mathbf{\tilde{r}}_{1}}\Binv\A\ket{\mathbf{\tilde{r}}_{1}} \\ - 
	& 2\bra{\mathbf{\tilde{r}}_{1}}\Binv\ket{\ri} \\ +
	&\notag 
	\bra{\mathbf{r}_{1M}} \Binv\A \ket{\mathbf{r}_{1M}} \\- 
	& 2\bra{\mathbf{r}_{1M}} \Binv \ket{\ri}]) d^{2}\mathbf{\tilde{r}}_{1}. &\notag
\end{align}

Eq. \ref{eq:collins_deriv1} is identical to the original Collins integral (Eq. \ref{eq:reduced_collins}), with the addition of two phase terms (shown in Eq. \ref{eq:phase_misalign_r2}) that are not functions of the variable of integration ($\tilde{\mathbf{r}}_{1}$). 

\begin{equation}
	\Phi_{misalign} = exp(-\frac{ik}{2}[\bra{\mathbf{r}_{1M}} \Binv\A \ket{\mathbf{r}_{1M}} - 2\bra{\mathbf{r}_{1M}} \Binv \ket{\ri}]])
	\label{eq:phase_misalign_r2}
\end{equation}

Factoring these terms out of the integrand reveals that we have arrived at an expression similar to Weber, where the field at the evaluation plane is proportional to the aligned Collins integral solution.
However, the result (shown in Eq. \ref{eq:collins_factored}) is instead expressed in terms of the common evaluation plane ($\mathbf{r}_{2}$) orthogonal to the optical axis.

\begin{align} \label{eq:collins_factored}
	E_{2}(\mathbf{r}_{2}) = \Phi_{misaligned}( K \iint_{-\infty}^{\infty} E(\mathbf{\tilde{r}}_{1}) exp(\frac{-ik}{2}[\bra{\ri}\D\Binv\ket{\ri} + &\notag \\
	\bra{\mathbf{\tilde{r}}_{1}}\Binv\A\ket{\mathbf{\tilde{r}}_{1}} - \\
	2\bra{\mathbf{\tilde{r}}_{1}}\Binv\ket{\ri}]) d^{2}\mathbf{\tilde{r}}_{1}.), &\notag
\end{align}

\begin{equation*}
	E_{2}(\mathbf{r}_{2}) = \Phi_{misaligned}E_{2,aligned}(\mathbf{r}_{2}).
\end{equation*}

Using the relation in Eq. \ref{eq:collins_factored}, a single beamlet can be propagated from a location in the entrance pupil of an optical system to every point in the evaluation plane simultaneously. Compared to the point-evaluation approach, this reduces the number of propagations done in beamlet decomposition algorithms by a factor equal to the number of points simulated in the evaluation plane and will result in favorable performance improvements. Eq. \ref{eq:collins_factored} illustrates that the standard Collins integral is used to propagate the complex amplitude of the beamlet, and $\Phi_{misaligned}$ applies a phase factor that aligns the beamlet to the evaluation plane. 


\section{Validation of the Plane-evaluation Algorithm}
\label{sec:results}

The plane-evaluation algorithm described in Section \ref{sec:algorithm} was implemented in Poke, an open-source ray-based physical optics package for Python 3.8+ \cite{Ashcraft_2023_poke}. Poke was, in part, inspired by the desire to develop an open-source platform to explore the limits of beamlet decomposition for use in diffraction simulation. Poke is capable of saving ray data from commercial and open-source ray tracers and compiling it into a writeable \verb"Rayfront" object that serves as the interface for conducting beamlet decomposition experiments.

\subsection{Test against Fresnel Diffraction}
Weber's formulation of the Collins' integral was originally for paraxial systems that could be represented by a single ray transfer matrix. To validate our approach, we first test the proposed formulation against a paraxial system composed of a single lens that focuses a decentered Gaussian beam. This can be equivalently modeled by the Fresnel diffraction integral by invoking Eq. \ref{eq:fresnel} and setting $U_{1}(r_{1})$ to be a vertically shifted Gaussian beam. The plane-evaluation algorithm can model this by invoking the Gaussian solution to the Collins integral, given by Eq. \ref{eq:gauss_prop}

\begin{equation}
	E_{2}(\ri) = \frac{e^{-ikl_{o}}}{\sqrt{|\A +\B\mathbf{Q}_{1}^{-1}}|} e^{-\frac{ik}{2}\bra{\ri}\mathbf{Q}_{2}^{-1}\ket{\ri}}
	\label{eq:gauss_prop}
\end{equation}
Where $\mathbf{Q}_{2}^{-1}$ is given by the well-known propagation law of the complex curvature matrix\cite{Siegman_1986}.
\begin{equation}
	\mathbf{Q}_{2}^{-1} = (\C + \D\mathbf{Q}_{1}^{-1})(\A + \B\mathbf{Q}_{1}^{-1})^{-1}
\end{equation}

The ray transfer matrix of interest is given by matrix multiplication of the ray transfer matrices that describe the refraction of a ray by a paraxial lens of focal length $f$ and propagating a distance equal to $f + \delta x$ and is given by Equation \ref{eq:abcd_sys}

\begin{equation}
	\begin{bmatrix}
		\A & \B \\
		\C & \D \\
	\end{bmatrix}
	=
	\begin{bmatrix}
		1 & 0 & f + \delta f & 0 \\
		0 & 1 & 0 & f + \delta f  \\
		0 & 0 & 1 & 0 \\
		0 & 0 & 0 & 1 \\
	\end{bmatrix}
	\begin{bmatrix}
		1 & 0 & 0 & 0 \\
		0 & 1 & 0 & 0 \\
		-1/f & 0 & 1 & 0 \\
		0 & -1/f & 0 & 1 \\
	\end{bmatrix}.
	\label{eq:abcd_sys}
\end{equation}

Here, we set up our lens to have a 12.7mm radius (half of the clear aperture) and a focal length of 100mm. The Gaussian beam is given a beam waist radius of 3mm, a wavelength of $1\mu m$, and shifted 10mm above the optical axis in the entrance pupil. We then propagate the Gaussian beam at a distance of 101mm so that it is outside of the focal region. 

\begin{figure}
	\centering
	\includegraphics[width=\textwidth]{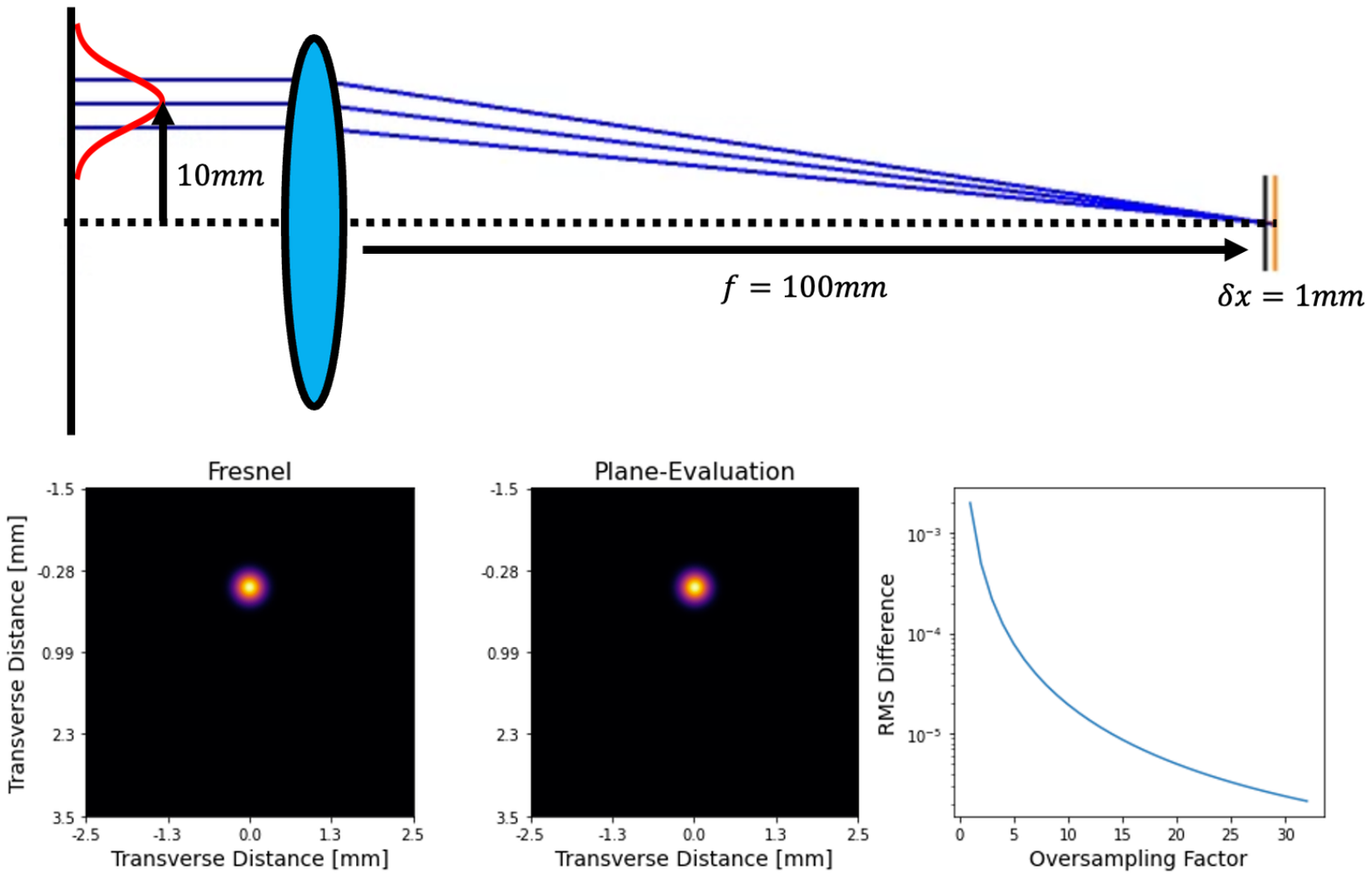}
	\caption{Consistency test of traditional Fresnel diffraction (left) and the proposed plane-evaluation algorithm (middle) for the case of a decentered Gaussian beam propagating to 1mm outside of focus. The Fresnel diffraction simulation on the left is plotted with an oversampling of 2. The results show that the irradiance profiles are nearly identical. Shown on the right is the RMS difference of the plots on the left and middle as a function of the oversampling of the Fresnel simulation. We see that we asymptotically approach zero as a function of oversampling, indicating that our proposed algorithm can correctly propagate a shifted Gaussian beam through an ABCD optical system.}
	\label{fig:fresnel_test}
\end{figure}

Shown in Fig. \ref{fig:fresnel_test} are the results of this simulation. Upon inspection, the Gaussian beams appear to be identical. When we compare the RMS difference of the two irradiances as a function of the oversampling factor (how much the Fresnel simulation was zero-padded), we observe that the difference asymptotically approaches zero. This is a strong indicator that we have derived the physics describing the propagation of a decentered Gaussian beam and can move forward to test their use in beamlet decomposition simulations.

\subsection{Test against Point-evaluation Algorithm}
In a prior study, we published the point-evaluation method of GBD to evaluate its suitability in simulations for astronomical high-contrast imaging. The methods are published in their entirety in Ashcraft et al. 2023 \cite{ashcraft_inreview} but are also a part of the Poke Python package in which we developed the plane-evaluation algorithm for this work. Using Poke as a simulation platform, we are able to easily compare the two implementations of GBD. We begin by testing both the point-evaluation and plane-evaluation algorithms against an analytical solution to the far-field diffraction pattern from a circular aperture: the Airy disk. We perform this simulation on a Ritchey-Chretien telescope with a 2.4m primary mirror modeled after the Hubble Space Telescope (HST) but remove the obscuration by the spiders and secondary mirror so that we can compare against the analytical solution. The prescription for this observatory is given in \hyperref[sec:appendixB]{Appendix B}. 

\begin{figure}
	\centering
	\includegraphics[width=\textwidth]{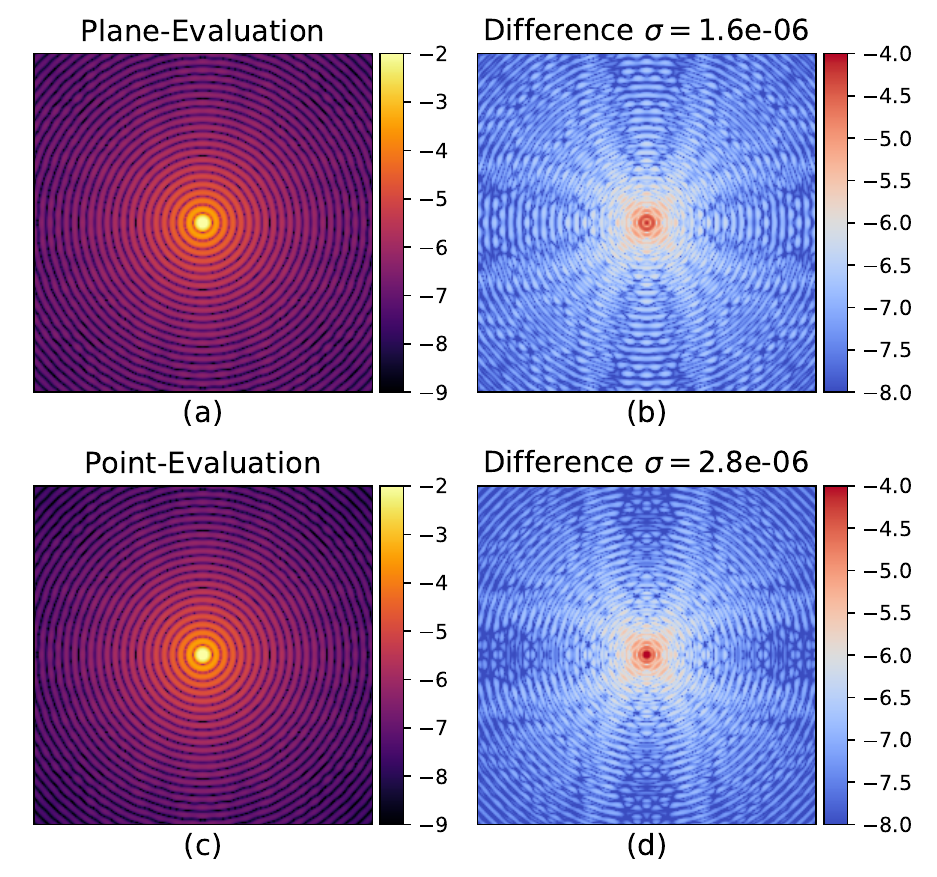}
	\caption{Comparison of the GBD PSFs (left) with their absolute differences with the analytical Airy function (right). The plane-evaluation algorithm produced the result in (a), and its absolute difference is in (b). The point-evaluation algorithm developed in \cite{ashcraft_inreview} is shown in (c), and its absolute difference is in (d). All data are $Log_{10}$ scaled to highlight the faint structure in the PSF. The high-frequency PSF residuals are very similar and of low magnitude, but the plane-evaluation algorithm shows smaller residuals, particularly in the PSF core, than the point-evaluation result. The RMS ($\sigma$) of the differences are given in the titles on the right, where the plane-evaluation algorithm has a lower RMS difference by about a factor of 2. The runtimes of the plane-evaluation and point-evaluation simulations were 113.78s and 1309.11s, respectively, using a computer with a 3GHz CPU and 16GB of RAM.}
	\label{fig:airy_compare}
\end{figure}

The analytical Airy pattern was simulated using the \textbf{POPPY} (Physical Optics Propagation in Python) optical propagation package \cite{Perrin12} using the \verb|poppy.misc.airy_2D| function. Shown in Fig. \ref{fig:airy_compare} are the results of computing the difference of the GBD PSFs with the analytical Airy function. The residuals shown in the right column are very similar, but note that the plane-evaluation algorithm actually reconstructs the core of the PSF slightly better and has a lower RMS error to the Airy function. We suspect that the decrease in RMS error is due to the fewer propagations required by the plane-evaluation approach. GBD is very sensitive to the differential computation of the ray transfer matrix, and the point-evaluation method computes a different ray transfer matrix for every evaluated point. The cumulative error from using differential ray tracing to propagate to each evaluated point likely results in a less accurate simulation. This is a strong indicator of the proposed algorithm's suitability to be used in beamlet decomposition simulations, but the biggest improvement occurs in the runtime of the simulations. The proposed algorithm was computed $11.5$ times faster than the point-evaluation method, meaning that we've gained an order of magnitude reduction in computation time in a PSF simulation that achieves twice the accuracy. 

\subsection{Aberrating the PSF}
\begin{figure}[H]
	\centering
	\includegraphics[width=\textwidth,trim={0cm 0cm 0cm 0}]{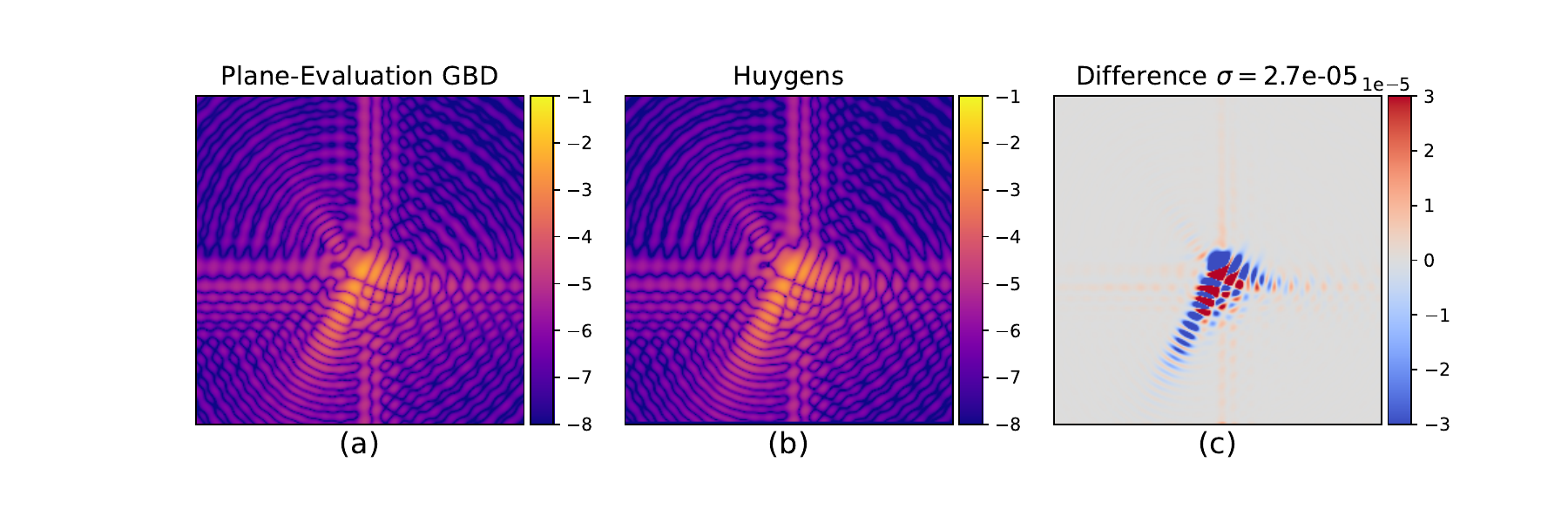}
	\caption{Comparison of the plane-evaluation algorithm's response to an aberrated optical system (a) to the Zemax OpticStudio Huygens PSF simulation (b) and their difference (c). (a) and (b) are plotted on a log scale to better resolve the PSF structure. We apply an aspheric sag to the primary mirror of the HST model described by Noll Zernike polynomials ($\Phi = \alpha Z_{7} + \alpha Z_{10}$, where $\alpha=10^{-7}$) to maintain the best focus position of the PSFs. Upon inspection, the fields in (a) and (b) are nearly identical, with an RMS difference of $\sigma = 2.7e^{-5}$. The largest features in (c) are on the order of 0.3$\%$ difference. They can be explained by the discrepancies between GBD and more exact scalar diffraction methods rather than inherent inaccuracies in the proposed algorithm.}
	\label{fig:huypsfcompare}
\end{figure}

For a final test of the functionality of the proposed algorithm, we simulated the aberration of the GBD PSF through surface errors on the primary mirror of the Hubble Space Telescope. We added an aperture with secondary support spiders and an aspheric sag to the primary mirror to aberrate the PSF and compare the plane-evaluation algorithm with that produced by Zemax's Huygens PSF simulation. The results are shown in Fig. \ref{fig:huypsfcompare}.

This tests the plane-evaluation algorithm's ability to recreate the diffraction effects from both sharp-edges and scalar aberration, indicating that, indeed, the proposed algorithm is suitable for imaging simulations while being much faster than the point-evaluation algorithm. We can now move forward to compare the speed of the plane-evaluation algorithm versus the point-evaluation algorithm to empirically examine how the proposed algorithm's efficiency scales on Central Processing Units (CPUs) and Graphical Processing Units (GPUs).


\section{Runtime Comparison}
\label{sec:runtime_compare}

Developing propagation routines with widely-used languages like Python enables us to easily scale up our computational power by using GPUs and high-performance computing resources. This is particularly useful in optical system modeling for rapid optimization and tolerancing. In this Section we report the key result of this study, the decrease in runtime achievable through use of the plane-evaluation algorithm we derived in Section \ref{sec:algorithm}. In Table \ref{tab:compute_resources}, we report on the two machines used to test the runtime of the beamlet decomposition algorithms. 

\begin{table}
	\centering
	\begin{tabular}{c c c c}
		\hline
		& TFLOPS & Cores & RAM  \\
		\hline
		Apple M1 & 0.154 & 8 & 16GB \\
		NVIDIA V100S & 8.2 & 5,120 & 32GB  \\
		\hline
		\hline
	\end{tabular}
	\caption{Description of the computational resources used in this investigation. The CPU results represent what an ``average" machine can do since it is performed on a consumer-grade laptop. The GPU results represent how the algorithms can scale to supercomputers. TFLOPS is teraFLOPS, or floating point operations per second, assuming double-precision floats.}
	\label{tab:compute_resources}
\end{table}

The Apple M1 processor is a consumer-grade CPU that is representative of what a typical researcher is likely to have available to them. The NVIDIA V100S GPU better represents how the proposed algorithm can scale to more expensive computational architectures. To best take advantage of these processing architectures, we construct the new beamlet decomposition method using two open-source Python packages.

\textbf{numexpr} is a Python package that allows for efficient multi-threading of elementary operations in Python and allows us to accelerate the remainder of the algorithm using CPUs. This package was used as the ``accelerated math" option for the \textbf{POPPY} optical propagation package\cite{Doug18} before GPUs were formally supported. On GPUs we anticipate even greater accelerations due to the parallelizability of the GBD algorithm. The problem can be broken up along each beamlet and each pixel, and distributed over the thousands of cores implemented in GPUs. The \textbf{cupy} Python package is a \textbf{numpy}-compatible library for scientific computing on NVIDIA GPUs. We integrated the interchangeable backend system of \textbf{prysm} (see Section 4A of Dube et al.\cite{Dube2022}) into Poke for easy migration of our calculations onto GPUs. 

\begin{figure}
	\centering
	\includegraphics[width=\textwidth]{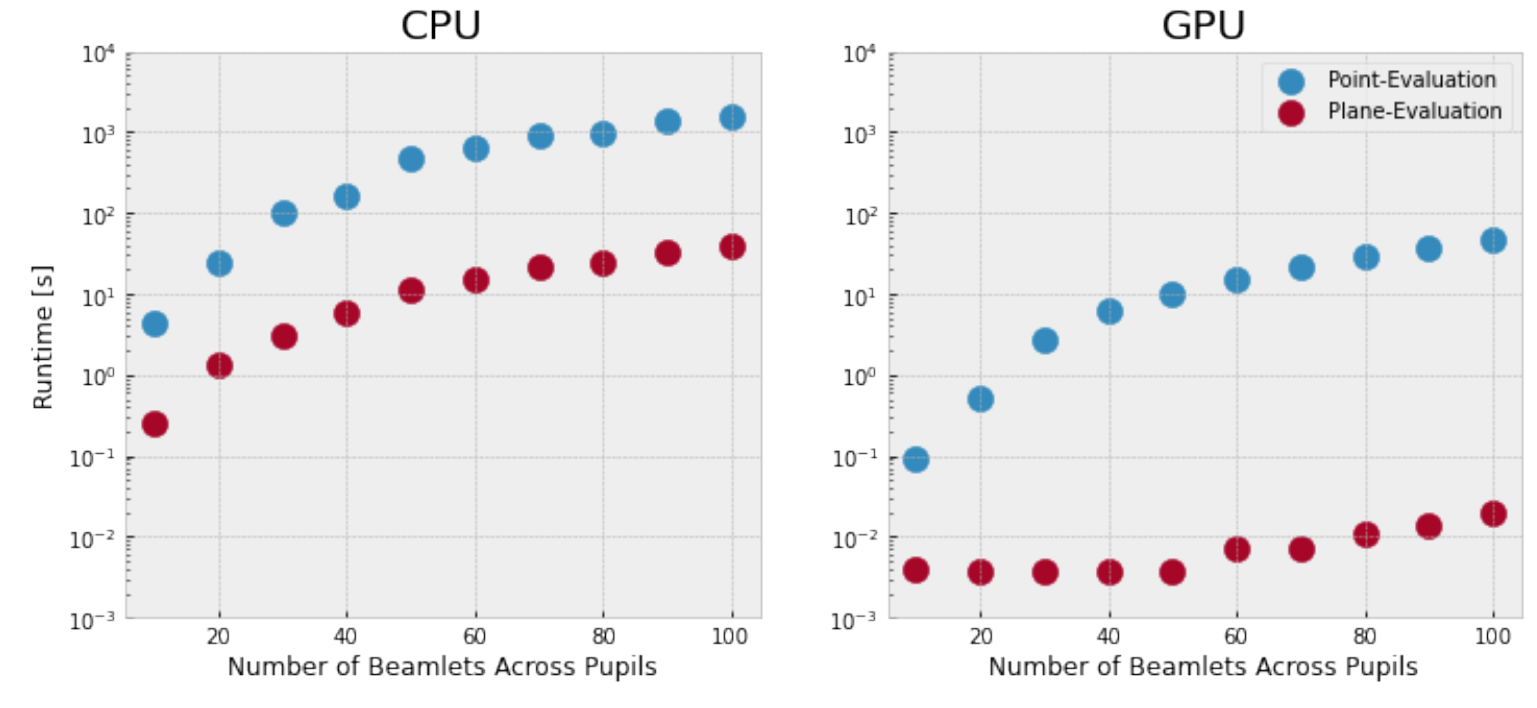}
	\caption{Runtime comparison of the point-evaluation (blue) and plane-evaluation (red) beamlet decomposition algorithms and their runtimes on the CPU (left) and GPU (right) listed in Table \ref{tab:compute_resources}. Each point represents a simulation where we compute the PSF of a telescope with a circular aperture with an entrance pupil diameter of 2.4m, a $\lambda = 551nm$, and a focal length of 57.6m on a grid of 256 x 256 pixels with a pixel scale of 2.8mas. The points are located at the runtime given a number of beamlets across the pupil. On the CPU, the mean runtime of the plane-evaluation algorithm is 34 times faster. On the GPU, the point-evaluation algorithm is 39 times faster, and the plane-evaluation algorithm is 1,760 times faster. The advancement made by this study is given by the relative runtime of the plane-evaluation algorithm on the GPU, which is 67,513 times faster than the point-evaluation algorithm on CPUs.} 
	\label{fig:runtime_total}
\end{figure}

The resulting algorithms are available on the Poke repository\cite{Ashcraft_2023_poke}, which are vectorized to maximize computational efficiency. In \hyperref[sec:appendixC]{Appendix C}, we also review the accelerations that were made to the elementary linear algebra functions that improved the runtime by 100x on CPUs. The results of our runtime comparison study are shown in Fig. \ref{fig:runtime_total}.

Across all cases studied, the plane-evaluation algorithm is at least one order of magnitude faster than the point-evaluation method. This is in part due to the analytical propagation of each beamlet to every point on the evaluation plane. A large portion of the computational resources of the point-evaluation method is spent determining the propagation of a beamlet to each point on the evaluation plane. Representing this information as a phasor instead allows for the rapid propagation of Gaussian beams. This also improves the memory efficiency because less information needs to be held by the RAM per beamlet. Less RAM per beamlet means that more can be processed at any given time, which is advantageous for our vectorized approach to GBD. The total runtime improvement contributed by this study using the data from Fig. \ref{fig:runtime_total} is the relative runtime decrease between the point-evaluation algorithm on CPUs (left blue) and the plane-evaluation algorithm on GPUs (right red). The mean runtime improvement per point comes out to about 67,513 times faster (not including the accelerations discussed in \hyperref[sec:appendixC]{Appendix C}). Another interesting result is that the runtime of the plane-evaluation algorithm on commercial CPUs is very close to the point-evaluation algorithm on a high-performance computing center GPU. This comparison highlights the contributions of this study: we've constructed an expression of GBD that is more accessible to the average researcher. These algorithmic advances in an open-source environment make GBD an accessible physical optics propagation technique, which we can use to learn more about degrees of freedom in a GBD simulation.


\section{Optimal Beamlet Parameter Search}
\label{sec:paramsearch}
\begin{figure}
	\centering
	\includegraphics[width=0.8\textwidth]{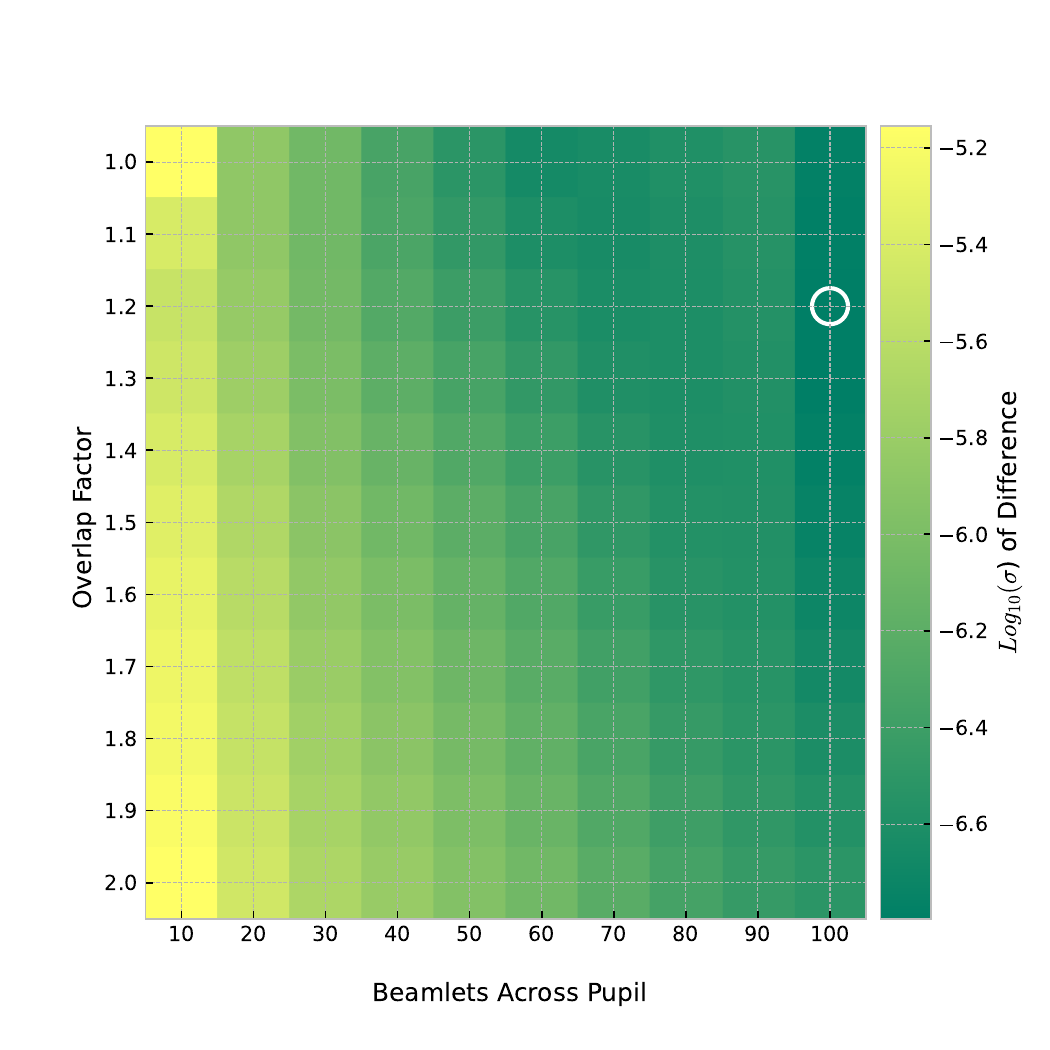}
	\caption{The $Log_{10}$ RMS difference to the analytical Airy function for a circular aperture as a function of the number of beamlets traced across the pupil and the overlap factor used. The white circle indicates the minimum error case for this parameter space. The total runtime to compute these PSFs on GPUs is 24 seconds.}
	\label{fig:optimal_param_search}
\end{figure}

One area of research that is not formally addressed in the literature is the best spatial distribution to decompose a circular aperture with Gaussian Beams. Fundamental Gaussian modes do not represent a complete set, so an analytical decomposition of a plane wave truncated by a circular aperture is not derivable. Harvey et al.\cite{Harvey15} introduced the Overlap Factor (OF) as a parameter for adjusting the width of evenly-spaced Gaussian beams. The OF is given in Eq. \ref{eq:OF}

\begin{equation}
	OF = \frac{N_{g} 2 \omega_{o}}{W},
	\label{eq:OF}
\end{equation}

where $N_{g}$ is the number of beamlets across an aperture, and $W$ is the width of the aperture. The goal is to determine the beamlet waist $\omega_{o}$, which is a function of the number of beamlets and the overlap factor. The optimal solution of these combinations is notionally suggested in Harvey et al. to be 1.5 (see Section 5 of \cite{Harvey15}), but an investigation into the tradeoffs between the number of beamlets and overlap factor was not explored for imaging systems. The proposed algorithm's efficiency simplifies this task, which we present in Fig. \ref{fig:optimal_param_search}.

In this experiment we evaluate the RMS difference of the analytical Airy function and a GBD simulation that aims to construct the Airy function. We evaluate this for overlap factors between 1 and 2 to explore the cases where adjacent beamlets are capable of smoothly reconstructing an aperture function. Note that extreme cases (e.g. $OF < 1$), beamlets do not meaningfully overlap and would have a much larger bearing on the accuracy of this simulation. Note also that the actual values of the RMS difference strongly depend on the pixel scale of the array computed and the array size, so we are principally interested in finding the lowest relative value in this trade space. The parameters used for the simulations in this section are given in \hyperref[sec:appendixD]{Appendix D}.

For imaging applications, it is clear that the overlap factors in this range do not have the largest bearing on the accuracy of the simulation. More critical is the number of beamlets used in the simulation. The RMS of the difference also more heavily weighs GBD's ability to capture the structure of the PSF core. Another way of saying this is that the result in Fig. \ref{fig:optimal_param_search} shows off the best combination of $OF$ and number of beamlets to simulate a radiometrically correct PSF. To better understand how a number of beamlets and overlap factors affect different spatial frequencies, we can compute the RMS difference of different annular zones in the result.

\begin{figure}
	\centering
	\includegraphics[width=\textwidth]{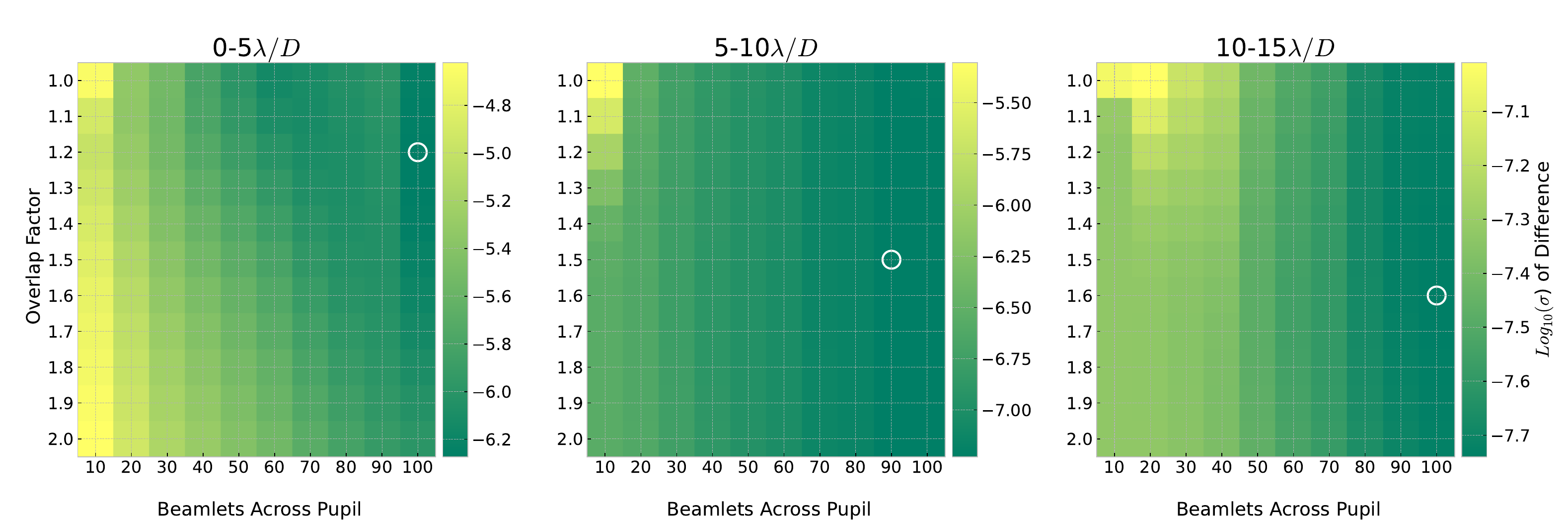}
	\caption{Comparison of the RMS difference between the simulated GBD PSFs and the analytical Airy function while varying overlap factor and number of beamlets across the pupil. (Left) The RMS difference of the PSFs from 0-5 $\lambda/D$. (Middle) The RMS difference of the PSFs from 5-10 $\lambda/D$. (Right) The RMS difference of the PSFs from 10-15 $\lambda/D$. The leftmost simulation follows the contour from  Figure \ref{fig:optimal_param_search}, but the higher-frequency information follows a different curve entirely. Generally, a PSF simulation is more accurate when more rays are used. However, in under-sampled cases, small overlap factors are better for low frequencies but worse for high frequencies. A white circle shows the minimum error case for a given parameter space.}
	\label{fig:fractional_param_search}
\end{figure}

In Fig. \ref{fig:fractional_param_search}, we compute the RMS difference of the simulated PSF with the analytical Airy function considering only certain regions in the PSF. We simulate the Airy function from 0 to 15 $\lambda/D$ for the model observatory and break it into three annular regions (0-5, 5-10, 10-15$\lambda/D$).

Fig. \ref{fig:fractional_param_search} illustrates a peculiarity of the degrees of freedom available in GBD. The overlap factor and number of beamlets impact different structures in the PSF differently. Low-spatial frequency information generally favors lower overlap factors, with a minimum of around 1.2. High-frequency information generally favors larger overlap factors, likely due to the mitigation of the amplitude ripple that is left from the beamlet decomposition \cite{Harvey15}. This example illustrates the necessity of more efficient algorithms for beamlet decomposition. We are able to rapidly explore the free parameters in the decomposition and determine that the optimal set of parameters is generally dependent on what structure in the PSF is critical in simulation. Of interest is that for the higher-frequency information, we've recovered the recommended overlap factor of 1.5 from Harvey et al \cite{Harvey15}.


\section{Discussion}
We present a new perspective on Weber's formulation of the general Collins integral that enables the expeditious computing of beamlet decomposition algorithms. The plane-evaluation method is tested against numerical and analytical results to verify its accuracy and is shown to outperform the point-evaluation method in both speed and accuracy. Moreover, our solution to the Collins integral maintains Weber's generality. The accelerated algorithm can be applied to the beamlet decomposition problem using any known solution to the Collins integral, including higher-order Gaussian beams and Worku and Gross' truncated and pulsed beam solutions. The algorithm is developed in the open-source Python package Poke, making it accessible to all investigators interested in developing beamlet decomposition as a viable physical optics propagation technique.

For the case of GBD, we determined that in general, more beamlets are critical to the accuracy of simulations and that the overlap factor has little bearing on the results once a sufficient number of beamlets are traced. However, this is solely for the case of propagating from a circular aperture to focus. Other applications, such as in non-imaging or illumination systems, may consider different figures of merit and degrees of freedom to optimize their beamlet distribution for maximum simulation accuracy.

The acceleration granted by the plane-evaluation algorithm enables the rapid computation of fields produced by GBD that is nearly as general as prior published methods. The only additional assumption imposed is that the points where we are evaluating the field fall on a plane ($\mathbf{r}_{2}$ in Fig. \ref{fig:weber_diagram}) that is orthogonal to the optical axis. Using the rapid evaluations enabled by this algorithm, we can use GBD propagation in optimization problems. We can also generate more highly sampled simulations in a shorter amount of time, which was one of the limitations of our prior study \cite{ashcraft_inreview}. 
The critical result of this study is that our derived propagation technique makes highly-sampled GBD practical. A reasonably highly-sampled simulation can be conducted in under a minute. Should the user have access to a GPU, they can take advantage of the parallelizability of GBD for even more accelerated simulations. We develop this algorithm in the Poke open-source Python package\cite{Ashcraft_2023_poke} so that it is freely available to all interested in pursuing the development of GBD as a propagation technique.

\subsection{Future Work}
We present a preliminary exploration into the ``optimal'' combination of a number of beamlets and overlap factor for simulating the PSF of a circular aperture. However, Worku and Gross have already proven that the accuracy of a simulation can be increased by employing truncated Gaussian beams to simulate aperture edges \cite{Worku19}. Performing a study similar to that of Section \ref{sec:results} that considers the distribution and number of truncated beams would be an optimal next step for using this algorithm in imaging simulations.

Non-imaging systems can also benefit from the proposed propagation method. High-energy laser systems, for example, require a precise understanding of the irradiance distribution of a given laser beam. With the proposed propagation technique, we can conduct highly sampled and rapid simulations of a higher-order Gaussian beam through an optical system without imposing the paraxial assumption on the system. 

\section{Appendix A: Summary of Weber's Formulation of the Collins Integral for Misaligned Optical Elements}
\label{sec:appendixA}
In this work, we depart from Weber's reformulation of the Collins Integral to arrive at the expression of the field common to several beamlets propagating through a given optical system. For a review of Weber's initial derivation, we briefly reproduce the results of Section 2 of their paper \cite{Weber06} to illustrate the differences.

Weber continues from where we left off in Eq. \ref{eq:collins_misalign} by also substituting the coordinate $\ri$ with the sum of the coordinate on the transversal plane $\mathbf{\tilde{r}_{2}}$ and the misalignment vector $\mathbf{r_{2M}}$. The result of the fully-expanded Collins integral for misaligned elements is given by Equation \ref{eq:collins_misalign_full}.

\begin{align} \label{eq:collins_misalign_full}
	E_{2}(\mathbf{\tilde{r}}_{2}) = & K \iint_{-\infty}^{\infty} E(\mathbf{\tilde{r}}_{1}) exp(\frac{-ik}{2}[\bra{\mathbf{\tilde{r}_{2}} + \mathbf{r_{2M}}}\D\Binv\ket{\mathbf{\tilde{r}_{2}} + \mathbf{r_{2M}}} &\notag
	\\ &+\bra{\rotilde}\Binv\A\ket{\rotilde}  \\
	& -2\bra{\rotilde}\Binv\ket{\mathbf{\tilde{r}_{2}} + \mathbf{r_{2M}}}]) d^{2}\mathbf{\tilde{r}}_{1} &\notag
\end{align}

Expanding the phasor in Eq. \ref{eq:collins_misalign_full} permits the separation of the terms that are linear in $\mathbf{\tilde{r}}_{1}$ and $\mathbf{\tilde{r}}_{2}$, as shown in Eq. \ref{eq:linear_terms},

\begin{align} \label{eq:linear_terms}
    E_{2}(\mathbf{\tilde{r}}_{2}) & = K \iint_{-\infty}^{\infty} E(\mathbf{\tilde{r}}_{1}) exp(-\frac{ik}{2}[ \notag \\
	& \bra{\mathbf{\tilde{r}}_{2}}\D\Binv\ket{\mathbf{\tilde{r}}_{2}} + 2\bra{\mathbf{\tilde{r}}_{2}}\Binv\ket{\mathbf{\tilde{r}}_{1}} + \bra{\mathbf{\tilde{r}}_{1}}\Binv\A\ket{\mathbf{\tilde{r}}_{1}} + \notag \\ 
    & \bra{\mathbf{r}_{1M}}\Binv\A\ket{\mathbf{r}_{1M}} + \bra{\mathbf{r}_{2M}}\D\Binv\ket{\mathbf{r}_{2M}} + 2\bra{\mathbf{r}_{2M}}\Binv\ket{\mathbf{r}_{1M}} +  \\
    & \bra{\mathbf{\tilde{r}}_{2}}\D\Binv\ket{\mathbf{r}_{2M}} + \bra{\mathbf{r}_{2M}}\D\Binv\ket{\mathbf{\tilde{r}}_{2}} + \bra{\mathbf{\tilde{r}}_{1}}\Binv\A\ket{\mathbf{r}_{1M}} + \notag \\
    & \bra{\mathbf{r}_{1M}}\Binv\A\ket{\mathbf{\tilde{r}}_{1}} + 2\bra{\mathbf{\tilde{r}}_{1}}\Binv\ket{\mathbf{r}_{2M}} + 2\bra{\mathbf{r}_{1M}}\Binv\ket{\mathbf{\tilde{r}}_{2}}])d^{2}\mathbf{\tilde{r}}_{1}, \notag
\end{align}

Where the top two rows of the phasor are the terms that aren't linear in $\mathbf{\tilde{r}}_{1}$ and $\mathbf{\tilde{r}}_{2}$ and the bottom two rows are the terms that are. Using the relation $\bra{\mathbf{r}_{1M}}\Binv\A\ket{\mathbf{\tilde{r}}_{1}} = \bra{\mathbf{\tilde{r}}_{1}}(\A\Binv)^{T}\ket{\mathbf{r}_{1M}}$ and the relations for symplectic matrices in Equation \ref{eq:symp1}-\ref{eq:symp5}, Weber shows that the terms linear in $\mathbf{\tilde{r}}_{1}$ and $\mathbf{\tilde{r}}_{2}$ vanish. What remains are the top two rows of the phasor in Equation \ref{eq:linear_terms}. The first row corresponds to the phasor of the aligned Collins integral (Eq. \ref{eq:reduced_collins}), and the phasor in the second row that isn't a function of the variable of integration. Consequently it can be factored out of the equation, resulting in the propagation law defined by Eq. \ref{eq:tilt_phase}

\begin{equation}
	E_{2}(\mathbf{\tilde{r}}_{2}) = E_{2}(\mathbf{\tilde{r}}_{2})_{align} exp(\frac{-ik}{2}[\bra{\mathbf{r}_{1M}}\Binv\A\ket{\mathbf{r}_{1M}} + \bra{\mathbf{r}_{2M}}\D\Binv\ket{\mathbf{r}_{2M}} - 2\bra{\mathbf{r}_{1M}}\Binv\ket{\mathbf{r}_{2M}}]),
	\label{eq:tilt_phase}
\end{equation}

where $E_{2}(\mathbf{\tilde{r}}_{2})_{align}$ is the aligned solution to the Collins integral in Eq. \ref{eq:reduced_collins}. Applying the propagation formulas given in Eqs. \ref{eq:ab_prop} and \ref{eq:cd_prop}, this can be simplified to Weber's original solution (Eq. \ref{eq:misalignphase}), which is the Collins integral expressed along the center of gravity vector that the beam propagates along. 

\begin{equation}
	\mathbf{r}_{2M} = \mathbf{A} \mathbf{r}_{1M} + \mathbf{B} \mathbf{\theta}_{1M},
	\label{eq:ab_prop}
\end{equation}
\begin{equation}
	\mathbf{\theta}_{2M} = \mathbf{C} \mathbf{r}_{1M} + \mathbf{D} \mathbf{\theta}_{1M}.
	\label{eq:cd_prop}
\end{equation}

\section{Appendix B: Hubble Space Telescope Model}
\label{sec:appendixB}

\begin{table}[H]
	\centering
	\begin{tabular}{c c c c c}
		\hline
		Surface & RoC [m] & Conic Constant & Distance [m] & Diameter [m] \\
		\hline
		M1 & -11.0400 & -1.00230 & -4.90607 & 2.40000 \\
		M2 & 1.35800 & -1.49686 &  6.40620 & 0.28112 \\
		\hline
		\\
	\end{tabular}
	\caption{Optical system prescription for the RC telescope based on the HST used in this investigation. All distances are given in meters. RoC stands for Radius of Curvature, and the sign convention is chosen such that negative values are concave and positive values are convex.}
	\label{tab:fiducial_observatory_specs}
\end{table}

\section{Appendix C: Notes on Elementary Function Computation}
\label{sec:appendixC}
GBD requires the evaluation of a determinant to compute the amplitude factor, a matrix inverse to compute the propagated $\mathbf{Q}_{2}^{-1}$, and the determination of eigenvalues to compute the Gouy phase $\mathbf{\Phi}_{gouy}$ \cite{Worku19}. In Python, one would typically call the \verb|numpy.linalg| library to compute these values. But after profiling our algorithm using the Viztracer\cite{Viztracer} Python profiler, we found that the largest contributors to our runtime were these functions.

\begin{figure}[H]
	\centering
	\includegraphics[width=\textwidth]{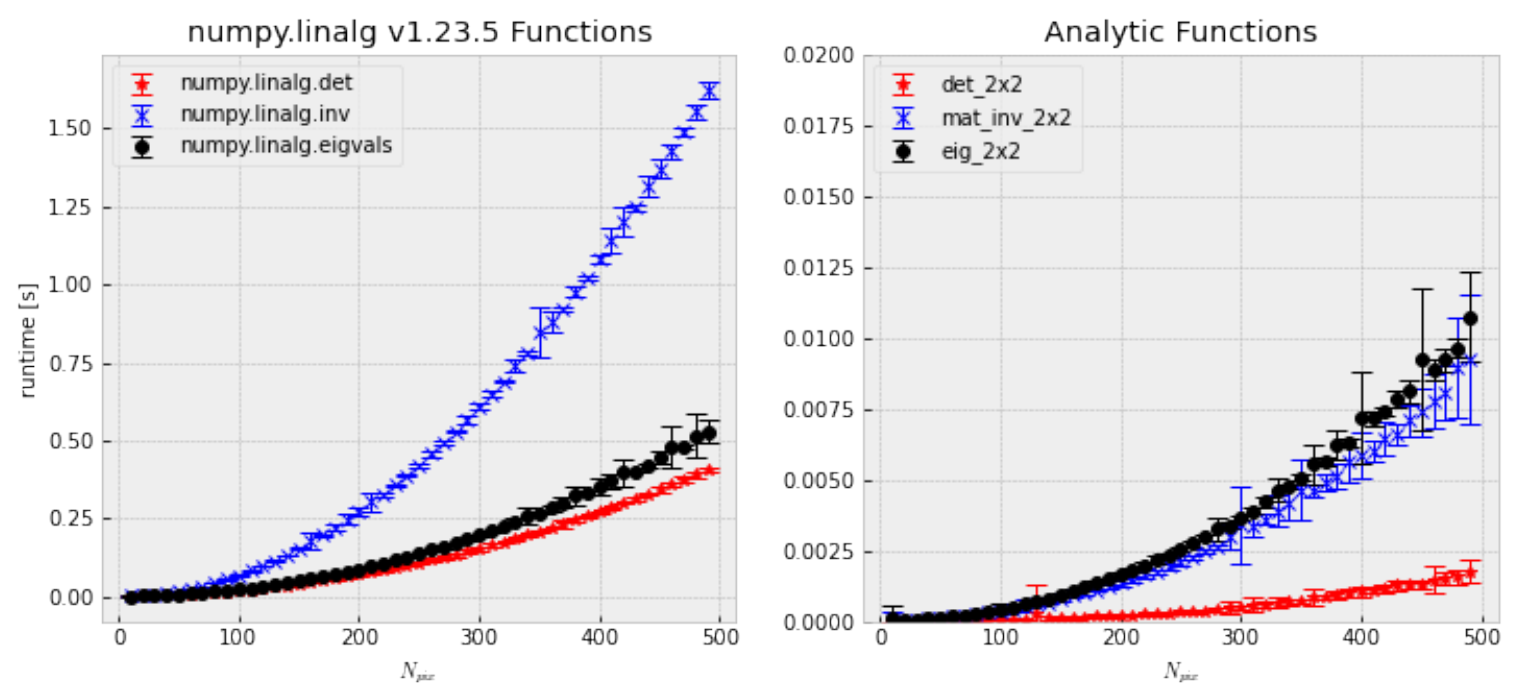} 
	\caption{Runtime comparison of elementary functions used in the GBD algorithm. Error bars represent the standard deviation of 25 trials on arrays of size $N_{pix} \times N_{pix} \times 2 \times 2$. \textbf{numpy}'s linear algebra library can operate on arrays of arbitrary shape, but the GBD algorithm only requires these operations on $2\times2$ matrices. For large arrays, our approach is $\approx$50 times faster for computing the determinant and 150-250 times faster for the inverse and eigenvalue computation on an M1 CPU.}
	\label{fig:elementary_funcs}
\end{figure}

Shown in Figure \ref{fig:elementary_funcs} are the results of a runtime comparison between these functions calling \verb|numpy.linalg| and our implementation of the same functions for 2x2 matrices, which can be found in college-level linear algebra textbooks. We only need these operations for 2x2 matrices, so they were trivial to implement in our algorithm. This shows that for large arrays (e.g. 500x500x2x2 
complex-valued matrices), we can improve the determinant calculation by a factor of ~50x and the inverse and eigenvalue calculations by a factor of 100-200x. These data were generated using \textbf{numpy} version 1.23.5.

\section{Appendix D: PSF Simulation Parameters for the Optimal Beamlet Parameter Search}
\label{sec:appendixD}

\begin{table}[H]
	\centering
	\begin{tabular}{c|c}
		\hline
		Parameter & Value  \\
		\hline
		Wavelength & 551nm \\
		EPD & 2.4m \\ 
		EFL & 57.6m \\
		npix & 512 $\times$ 512\\
		pixelscale & 2.8mas \\
		\hline
		\hline
	\end{tabular}
	\caption{Parameters used in the optimal beamlet parameter search. These data were used with the Ansys Zemax OpticStudio ray tracing engine and the Poke open-source physical optics package. They can be found in the experiments/weber directory of the Poke GitHub repository.}
	\label{tab:my_label}
\end{table}

\section{Acknowledgements}
This research made use of several open-source Python packages, including \textbf{POPPY}\cite{Perrin12}, \textbf{HCIPy}\cite{por2018hcipy}, \textbf{prysm}\cite{Dube2019}, \textbf{numpy}\cite{harris2020array}, \textbf{matplotlib}\cite{Hunter:2007}, \textbf{ipython}\cite{PER-GRA:2007}, \textbf{scipy}\cite{2020SciPy-NMeth}, and \textbf{numexpr}\cite{robert_mcleod_2018_2483274}. This research made use of High Performance Computing (HPC) resources supported by the University of Arizona TRIF (Technology and Research Initiative Fund), UITS, and Research, Innovation, and Impact (RII) and maintained by the UArizona Research Technologies department. This work was supported by a NASA Space Technology Graduate Research Opportunity.

\section*{Data Availability}
Data underlying the results presented in this paper are available at the Poke repository on GitHub \url{github.com/Jashcraf/poke} \cite{Ashcraft_2023_poke,jaren_ashcraft_2023_7951685}.

\section*{Disclosures} 
The authors declare no conflicts of interest.


\bibliography{sample}

\end{document}